\documentclass[12pt]{iopart}
\usepackage{amssymb,wasysym}
\usepackage{graphicx}
\usepackage{subfig}

\begin{document}
\title{How strong is the evidence for accelerated expansion?}
\author{Marina Seikel$^1$ and Dominik J Schwarz$^2$}
\address{Fakult\"at f\"ur Physik, Universit\"at Bielefeld, Postfach
  100131, 33501 Bielefeld, Germany}
\ead{\mailto{$^1$mseikel@physik.uni-bielefeld.de},
  \mailto{$^2$dschwarz@physik.uni-bielefeld.de}} 

\begin{abstract}
We test the present expansion of the universe
using supernova type Ia data without making any assumptions about the
matter and energy content of the universe or about the
parameterization of the deceleration parameter. We assume the
cosmological principle to apply in a strict sense. The result 
strongly depends on the data set, the light-curve fitting
method and the calibration of the absolute magnitude used for the
test, indicating strong systematic errors. Nevertheless, in a
spatially flat universe there is at least a 5$\sigma$ evidence for
acceleration which drops to 1.8$\sigma$ in an open universe. 
\end{abstract}


\section{Introduction}
The first evidence for an accelerated expansion of the universe at the
present epoch was presented in the late 1990s by Riess et
al.~\cite{riess98} and Perlmutter et al.~\cite{perlmutter}.
Fitting a cosmological model including a cosmological constant matched
the supernovae type Ia of the considered data sets
significantly better than a model without such a constant. 
In the following years, a lot of effort was put into getting larger
supernova (SN) sets of increasing quality \cite{jha,snls,gold07,essence}. 
On the other hand, a variety of
cosmological models have been developed that could be fit to the
newest data sets. Those models are characterized by certain parameters
which are used to derive a redshift-luminosity relation that can
be compared to the observed values.
The problem with this so-called dynamical
approach is that it is impossible to test without assumptions on
the matter and energy content of the universe
if there really is a phase of acceleration in the expansion history of
the universe.

Some authors tried to avoid this
problem by taking a kinematical approach, i.e. they only considered
the scale factor $a$ and its derivatives, such as
the deceleration parameter $q(z)$, without using
any model specific density parameters or a dark energy equation of
state. The first kinematical analysis of SN data was done by
Turner and Riess \cite{turner} who considered averaged values $q_1$ for
redshift $z<z_1$ and $q_2$ for $z>z_1$ concluding that a present acceleration
and a past deceleration is favoured by the data. Other authors tested
a variety of special parameterizations of $q(z)$
\cite{gold04,elgaroy} or used  $a(t)$ \cite{wang} or the Hubble rate $H(z)$
\cite{john}. Instead of considering a special parameterization, 
a more general approach has been done by Shapiro and Turner
\cite{shapiro} who expanded the deceleration parameter $q$ into
principal components. 
Rapetti et al.~\cite{rapetti} expanded the jerk parameter $j$ into a
series of orthonormal functions.
However, one has to be careful when doing a series expansion in $z$ as
SNe with a redshift larger than 1 are not within the radius of
convergence. This problem can be solved by reparameterizing redshift
\cite{cattoen}. 

In the present work we will neither make any assumptions about the content
of the universe, nor about the parameterization of $q(z)$ or another
kinematical quantity. Moreover, we do not need to assume the validity
of Einstein's equations.
Instead, we will only ask the question if the
hypothesis holds that the universe never expanded accelerated 
between the time when the light was emitted from a
SN and today. Our assumption for the test is that the universe is  
isotropic and homogeneous. This is certainly not true for the real
universe in a strict sense. However, we follow the standard approach
in assuming that the cosmic structure does not modify the observed SN
magnitudes and redshifts apart from random peculiar motion.
The basic idea of this analysis has already been
presented by Visser \cite{visser} and has been applied to SN data by
different groups \cite{santos, gong}. 
But while Santos et al.~\cite{santos} made some mistakes in their
analysis (that we will discuss later in section \ref{concl}) and
Gong et al.~\cite{gong} only state that
accelerated expansion is ``evident'', we are able to give a
quantitative value for this evidence. Additionally, we study the size
of systematic effects.

For the fit one usually marginalizes over a function of the absolute
magnitude $M$ and the Hubble constant $H_0$ because these two values
cannot be determined independently by only considering
SNe. Marginalization is not suitable for our analysis because in order
to do so a special cosmological model or at least a parametization of
$q(z)$ has to be inserted, which is what we want to avoid. But the SNe
can be calibrated by cepheid meassurements and thus the values of $M$
and $H_0$ are determined. Using this additional information, no
marginalization is needed. As there is still a controversy
\cite{jackson} about the 
appropriate calibration method, we will take two quite different
results of calibration into account \cite{riess, sandage}.

In order to test the robustness of our analysis, we consider two
different SN data sets (the 2007 Gold sample \cite{gold07} and the
ESSENCE set \cite{essence}), where the data of the ESSENCE set are
once obtained by the multicolour light-curve shape (MLCS2k2) \cite{jha}
fitting method and once by the spectral adaptive light-curve template
(SALT) \cite{guy}. Both calibration methods are applied to each 
set. The analysis shows that in all cases the data indicate an
accelerated expansion. But the confidence level at which acceleration
can be stated strongly depends on the data set, the fitting method and
the calibration. We begin our analysis by assuming a flat universe,
but later also consider the cases of an open and a closed universe.

\section{Method}
We want to keep our test as model-independent as possible, but still a
few assumptions have to be made. We consider inflationary cosmology to
be correct. This implies large scale homogeneity and isotropy of the
universe as well as spatial flatness. (Yet, we will give up the assumption
of a flat universe later in section \ref{openclosed}.)
We also assume that cosmological
averaging (here along the line of sight) does not modify the result
obtained in a Friedmann-Lema\^{i}tre model.
In such a universe, the
luminosity distance $d_{\mbox{\scriptsize L}}(z)$ is given by
\begin{equation}
d_{\mbox{\scriptsize L}}(z) = (1+z) \int_0^z \frac{\rmd
  \tilde{z}}{H(\tilde{z})}\;.
\end{equation}
As we are interested in the question whether the universe accelerates
or decelerates at the present epoch, we need to examine the
deceleration parameter $q(z)$: If $q$ is positive, the universe
expands decelerated, for a negative $q$ it accelerates.  $q(z)$ can be
expressed in terms of the Hubble parameter $H(z)$:
\begin{equation}
q(z) = \frac{H'(z)}{H(z)} (1+z) -1 \,,
\end{equation}
where the prime denotes the derivative with respect to $z$.
Integrating this equation yields
\begin{equation}
\ln \frac{H(z)}{H_0} =
\int_0^z\frac{1+q(\tilde{z})}{1+\tilde{z}}\,\rmd \tilde{z}\;. 
\end{equation}

Our null hypothesis is that the universe never expanded accelerated,
i.e. $q(z)\ge 0$ for all $z$. Under this assumption, the above
equation turns into the inequality
\begin{equation}
\ln \frac{H(z)}{H_0} \ge \int_0^z\frac{1}{1+\tilde{z}}\,\rmd \tilde{z}
= \ln (1+z) 
\end{equation}
or $H(z)\ge H_0(1+z)$. Thus, for the luminosity distance we have
\begin{equation}\label{dlinequ}
d_{\mbox{\scriptsize L}}(z) \le (1+z) \frac{1}{H_0} \int_0^z
\frac{\rmd \tilde{z}}{1+\tilde{z}} = (1+z)\frac{1}{H_0}\ln (1+z)\;.
\end{equation}

In order to test our hypothesis, we consider different data sets of
SNe type Ia. If the observed luminosity distance is
significantly larger than the luminosity distance of a universe with a
constant $q=0$, the hypothesis can be rejected at a high confidence
level. Note that the rejection of the hypothesis does not mean that
there was no deceleration between the time the light was emitted from
a SN until now, it only gives evidence that there was at some
time a phase of acceleration. Thus, we cannot determine the transition
redshift between deceleration and acceleration. This restriction to
our analysis comes from the integral over redshift in the calculation
of $d_{\mbox{\scriptsize L}}(z)$.

As a first step, the data in the considered sets have to be
calibrated consistently. The distance modulus $\mu$ is related
to the luminosity distance by
\begin{equation}
\mu=m-M = 5\log d_{\mbox{\scriptsize L}} +25 \,,
\end{equation}
where $d_{\mbox{\scriptsize L}}$ is given in units of Mpc. The
distance modulus of the SNe Ia in those sets is always given in
arbitrary units because the absolute magnitude $M$ cannot be measured
independently of the Hubble constant $H_0$. Only the apparent
magnitude $m$ and thus the relative distance moduli are measured at
high precision. In order to determine the absolute magnitude, the SNe
have to be calibrated by measuring the distance of cepheids in the
host galaxies. Then also the Hubble constant can be determined. But
there is still disagreement between different groups about the correct
calibration analysis. In this work we will consider two results for
$M$ and $H_0$, namely that of Riess et al.~\cite{riess} and that of
Sandage et al.~\cite{sandage}. In the following we will refer to those
results as Riess calibration and Sandage calibration,
respectively. The difference in the calibration comes mainly from the
different assumptions on the evolution of the cepheid
period-luminosity relation with host metallicity \cite{jackson}.
So for preparing the data for our
analysis, first the distance moduli $\mu_i$ have to be adjusted to the
assumed absolute magnitude.

We define the magnitude $\Delta\mu_i$ as the observed distance modulus
$\mu_i$ of 
the $i$-th SN minus the distance modulus in a universe with
constant deceleration parameter $q=0$ at the same redshift $z_i$:
\begin{equation}\label{deltamui}
\Delta\mu_i = \mu_i - \mu(q=0) = \mu_i - 5\log\left[
  \frac{1}{H_0}(1+z_i)\ln(1+z_i) \right] -25 \,.
\end{equation}
If the error in redshift and the peculiar velocities of the SNe are
already included in the error $\sigma_{\mu_i}$ given in a certain data
set, then the error $\sigma_i$ of $\Delta\mu_i$ equals
$\sigma_{\mu_i}$. Otherwise, the resulting error of $\Delta\mu_i$ is
calculated by:
\begin{equation}
\sigma_i = \left[\sigma_{\mu_i}^2 + \left( 5\frac{\ln(1+z_i)
    +1}{(1+z_i)\ln(1+z_i)\ln10} \right)^2 \left(\sigma_z^2 +
  \sigma_v^2 \right) \right]^{\frac{1}{2}} \,.
\end{equation}

Let $\mu_{\mbox{\scriptsize th}}(z)$ be the theoretical
distance modulus of a cosmological model, which describes the
expansion of the universe correctly. Then
\begin{equation}
\Delta\mu_{\mbox{\scriptsize th}}(z_i) = \mu_{\mbox{\scriptsize
    th}}(z_i) - \mu(q=0)
\end{equation}
would be the ``true'' value corresponding to the meassured SN value
$\Delta\mu_i$.  The null hypothesis for our test is that the universe
never expanded accelerated, i.e. $\Delta\mu_{\mbox{\scriptsize th}}
\le 0$ for each SN. We reject that hypothesis if the measured
value $\Delta\mu_i$ lies above a certain action limit
$A_{\mbox{\scriptsize a}}$, otherwise
we accept it. We want to keep the risk low that we conclude a late
time acceleration of the universe, when there is indeed no
acceleration at all. Therefore the action limit must be relatively
high. If we want the confidence level for concluding an accelerated
expansion to be 99\%, the action limit must be $A_{\mbox{\scriptsize
    a}}(99\%)=2.326\sigma_i$. For a confidence level of 95\%, the
limit is $A_{\mbox{\scriptsize a}}(99\%)=1.645\sigma_i$. Here we
assume that the measured values $\Delta\mu_i$ at a given redshift
follow a normal distribution.

On the other hand, we can test the hypothesis that the universe
expanded accelerated all the time from light emission of a SN
until today. This hypothesis can be rejected at a 99\% CL if
$\Delta\mu_i$ is below the action limit $A_{\mbox{\scriptsize
    d}}(99\%)=-2.326\sigma_i$. That would mean that there must have
been a phase of deceleration in the late time universe, but it would
not exclude a phase of acceleration.

\section{Data Sets}
The important parameters of the SNe are obtained by fitting
their light-curves. The results depend on the fitter that is used. The
most common fitter is MLCS2k2 \cite{jha}. Here we will also consider
the SALT fitting method \cite{guy}. The main difference between the
two methods is in how the peak luminosity corrections are determined
by the colour of the SNe \cite{conley}. While MLCS2k2 assumes that the
corrections that have to be made are only due to dust, SALT takes an
empirically approach to determine the relation between the colour and
the luminosity correction.

For the analysis we take the following SN Ia data sets:
\begin{description}
\item{Gold 2007 (MLCS2k2):} the Gold sample by Riess et al.~(2007)
  \cite{gold07}, which was obtained by using the MLCS2k2 fitting
  method
\item{ESSENCE (MLCS2k2):} the set given by Wood-Vasey et al.~(2007)
  \cite{essence}, which includes data from ESSENCE, SNLS \cite{snls}
  and nearby SNe \cite{jha}, fitted with MLCS2k2
\item{ESSENCE (SALT):} the same set fitted with SALT
\end{description}
As suggested by Riess et al.~\cite{gold07}, we discarded all SNe with
a redshift of 
$z<0.0233$ from the Gold sample, which leaves us 182 SNe with
$0.0233<z<1.755$. In ESSENCE
(MLCS2k2) and ESSENCE (SALT), the SNe with bad light curve fits were
rejected for each fitting method seperately. This leaves 162 SNe for
ESSENCE (MLCS2k2) and 178 for ESSENCE (SALT) with $0.015<z<1.01$. 153
of the SNe are
contained in both sets. Due to the differences of the SNe contained in
ESSENCE (MLCS2k2) and ESSENCE (SALT) we will refer to them as
different sets in the following. Nevertheless, you should keep in mind
that they share a large number of SNe and thus are not
independent sets.

The Riess calibration yields a
value for the V-band magnitude of $M_{\mbox{\scriptsize
    V}}(t_0)=-19.17\pm 0.07$mag, where $t_0$ is the time of the
B-band maximum. For the Gold 2007 set $M_{\mbox{\scriptsize V}}(t_0)$ was
considered to be $-19.44$mag. Therefore, in order to get the
appropriate distance modulus $\mu=m-M$, we need to substract 
$0.27$mag from the value given in the Gold sample. From the distance
modulus values in the ESSENCE sets $0.22$mag have to be substracted
because in these sets a B-band magnitude of $M_{\mbox{\scriptsize
B}}=-19.5$mag is assumed and Riess et al.~\cite{riess} give the relation
$M_{\mbox{\scriptsize B}}-M_{\mbox{\scriptsize V}}=-0.11$. For this 
calibration one gets a Hubble constant of $H_0=73\pm
4(\mbox{statistical}) \pm 5(\mbox{systematic})$.

The results of Sandage et al.~\cite{sandage} for the SNe absolute
magnitudes are $M_{\mbox{\scriptsize V}}=-19.46$mag and
$M_{\mbox{\scriptsize B}}=-19.49$mag and for the Hubble 
constant $H_0=62.3 \pm 1.3$(statistical) $\pm$
5.0(systematic). So considering the Sandage calibration, we have to
add 0.02mag to distance moduli of the Gold sample and substract
0.01mag from the values given in the ESSENCE sets.

\section{Results for a flat universe}

\subsection{Single SNe}
\begin{figure}
\centering \subfloat[Gold 2007 (MLCS2k2), Riess calibration]{
  \includegraphics*[width=7.5cm]{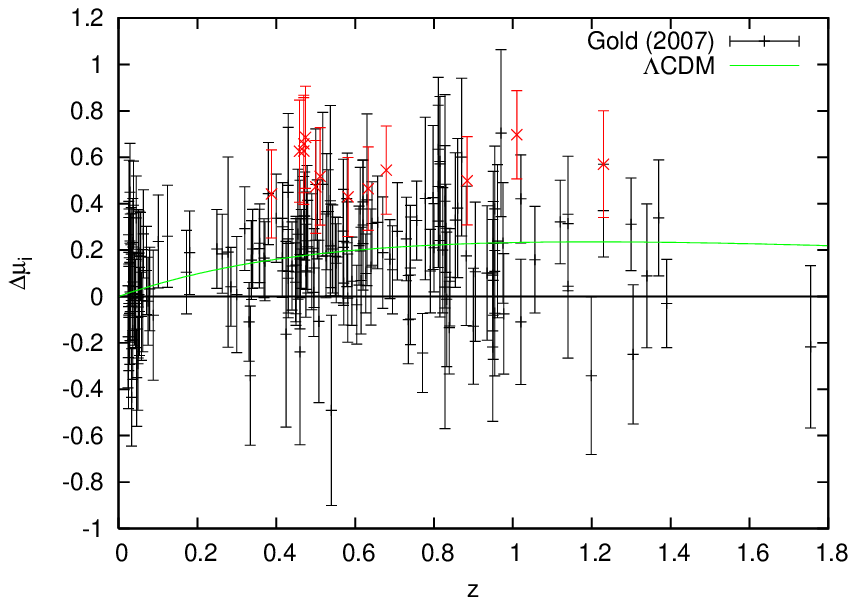}} \hfill
\subfloat[Gold 2007 (MLCS2k2), Sandage calibration]{
  \includegraphics*[width=7.5cm]{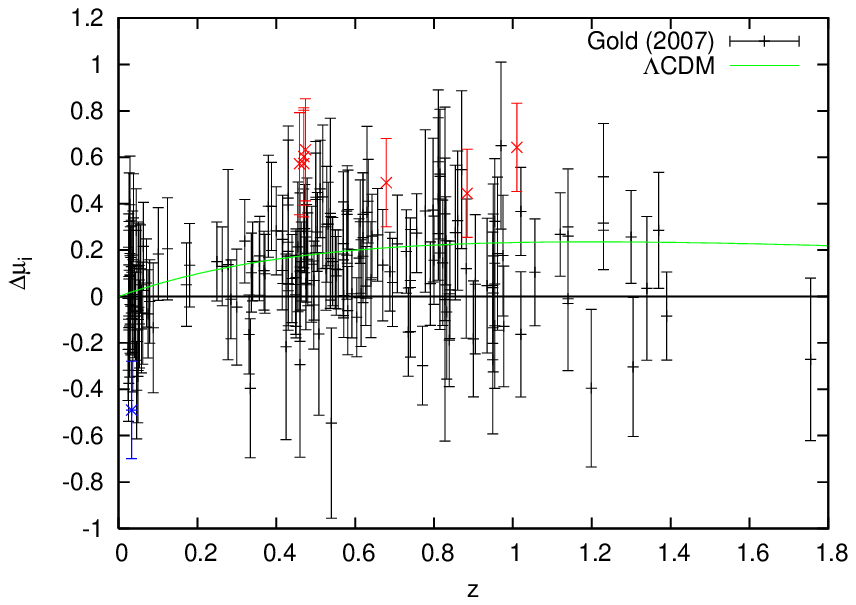}}
\\ \subfloat[ESSENCE (MLCS2k2), Riess calibration]{
  \includegraphics*[width=7.5cm]{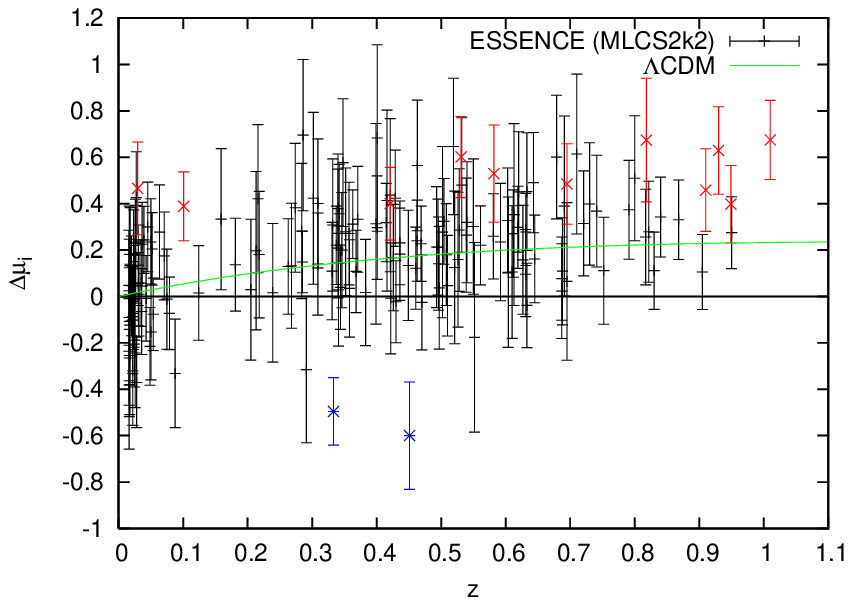}} \hfill
\subfloat[ESSENCE (MLCS2k2), Sandage calibration]{
  \includegraphics*[width=7.5cm]{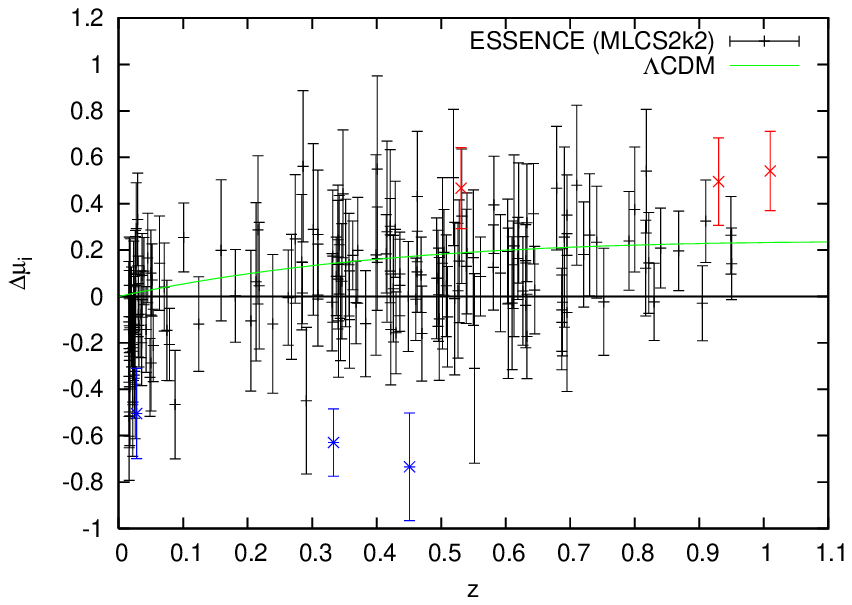}}
\\ \subfloat[ESSENCE (SALT), Riess calibration]{
  \includegraphics*[width=7.5cm]{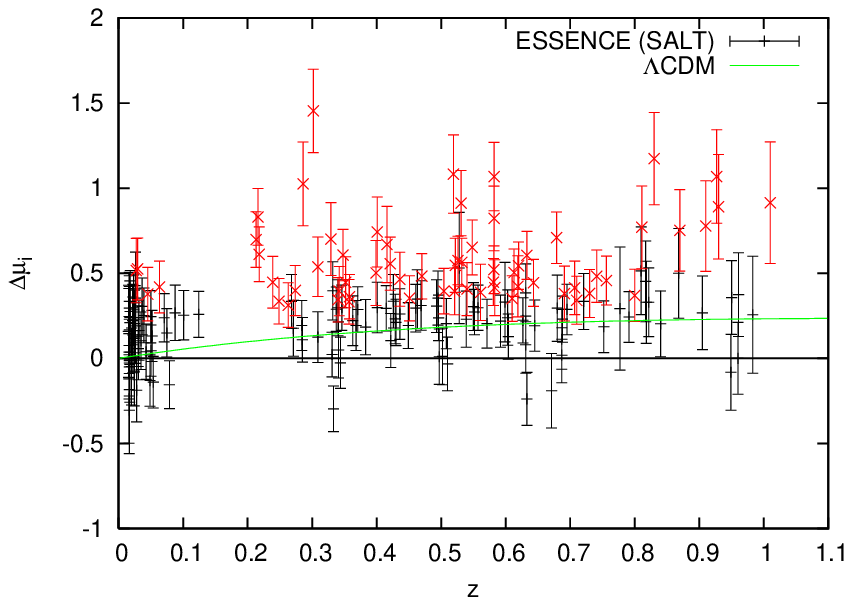}} \hfill
\subfloat[ESSENCE (SALT), Sandage calibration]{
  \includegraphics*[width=7.5cm]{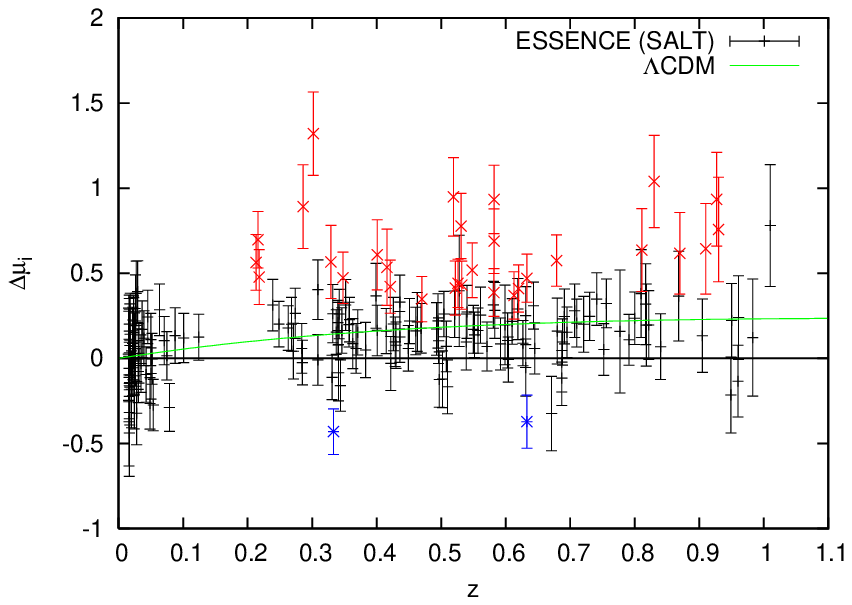}}
\caption{Magnitude $\Delta\mu_i$, as defined in equation
  \eref{deltamui}, for the three data sets and the two
  calibrations. SNe that indicate acceleration at 99\% CL are
  plotted in red ($\times$), those that indicate deceleration in blue
  ($\hexstar$). Also shown is the curve for a flat $\Lambda$CDM model
  with $\Omega_m=0.3$ (which is not a fit to the data). Note that the
  SALT method leads to a larger spread in $\Delta\mu_i$, whereas the
  Gold set extends to higher redshifts.} 
\label{deltamufig}
\end{figure}

The relevant parameter for our analysis is $\Delta\mu_i$ given by
equation \eref{deltamui}. Its values are plotted in figure
\ref{deltamufig} for the SNe of the three data sets and the two
different calibration methods. The curve for a flat $\Lambda$CDM model
(with $\Omega_{\mbox{\scriptsize m}}=0.3$) is shown in the plots
in order to give the reader a notion of how the redshift-dependency of
$\Delta\mu$ could look like.  Most of the data values are positive
which indicates an accelerated expansion. 
A similar diagram has been presented by Gong et al.~\cite{gong} for
single SNe as well as for binned SN data where
they used a combined Gold and ESSENCE data set. They kept the arbitrary
value of the absolute magnitude $M$ given in the set and then determined
the Hubble constant by using SNe with $z\le 0.1$ to be $H_0=66.04$. At
this point they stopped their analysis by concluding that due to the
large number of SNe that lie above the curve of a universe with
$q=0$, accelerated expansion is evident.
But the fact that there are
also several data points below that curve and that the errors
$\sigma_i$ (with a typical value of 0.23mag for MLCS2k2 and 0.18mag
for SALT) are
approximately of the same size as the values $\Delta\mu_i$ give rise
to the question how certain acceleration really is. Thus, we will
provide a more quantitative analysis in this work.

We counted the SNe of each set whose values of $\Delta\mu_i$
were above the action limit $A_{\mbox{\scriptsize
    a}}(95\%)=1.645\sigma_i$ and those above $A_{\mbox{\scriptsize
    a}}(99\%)=2.326\sigma_i$ (which indicates acceleration at a 95\%
and a 99\% CL, respectively) and those with values below
$A_{\mbox{\scriptsize d}}(95\%)=-1.645\sigma_i$ and below
$A_{\mbox{\scriptsize d}}(99\%)=-2.326\sigma_i$ (which indicates
deceleration). The results are shown in table \ref{acctab1}. The
clearest evidence for an accelerated expansion is given by the ESSENCE
(SALT) set with the Riess calibration with 64 SNe
indicating acceleration at a 99\% confidence level and none indicating
deceleration. Also in the other sets accelerated expansion is
preferred with the exception of ESSENCE (MLCS2k2) in the Sandage
calibration where none of the two expansion histories is preferred in
the test using the action limit $A(99\%)$.

\begin{table}
\begin{indented}
\lineup
\caption{\label{acctab1} Number of SNe indicating acceleration or
  deceleration at 95\% and 99\% CL for the different data
  sets and calibrations. Also given is the total number of SNe in each set. The most
  different results are highlighted.}
\item[]
\begin{tabular}{@{}lllllll} \br 
&\centre{2}{Gold 2007 (MLCS2k2)}
  &\centre{2}{ESSENCE (MLCS2k2)} &\centre{2}{ESSENCE (SALT)} \\ 
& Riess & Sandage & Riess & Sandage & Riess & Sandage \\\mr
acceleration (95\% CL) & \037 & \027 & \037 & \014 & \096 & \053\\ 
acceleration (99\% CL) & \013 &\0\07 & \011 &\0\0{\bf 3}&\0{\bf 64}&\030 
\\\mr 
deceleration (95\% CL) &\0\02 &\0\04 &\0\03 &\0\07 &\0\01 &\0\07\\ 
deceleration (99\% CL) &\0\00 &\0\01 &\0\02 &\0\0{\bf 3}&\0\0{\bf 0}&\0\02 
\\\mr 
number of SNe & 182 & 182 & 162 & 162 & 178 & 178 \\ \br
\end{tabular}
\end{indented}
\end{table}

\begin{figure}
\centering \subfloat[]{\label{mudiff}
  \includegraphics*[width=7.5cm]{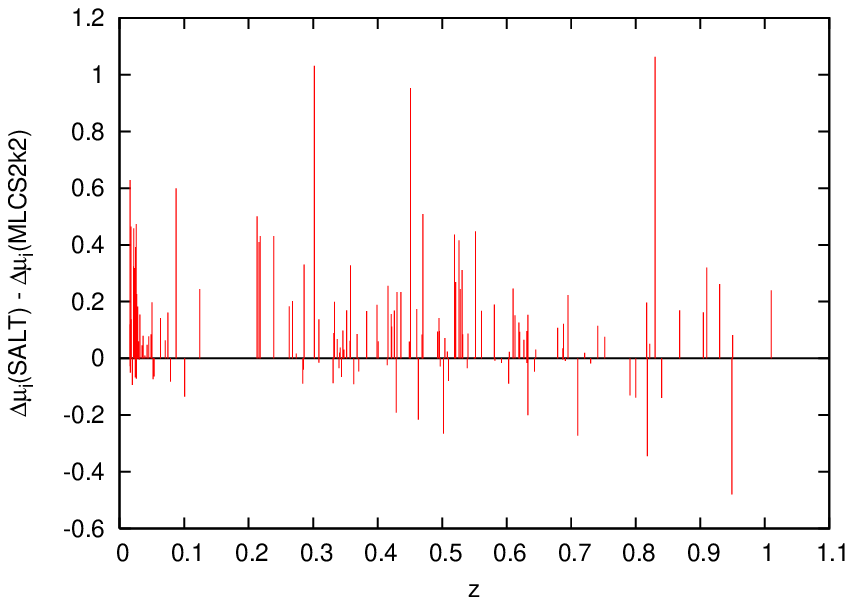}} \hfill
\subfloat[]{\label{weightmudiff}
  \includegraphics*[width=7.5cm]{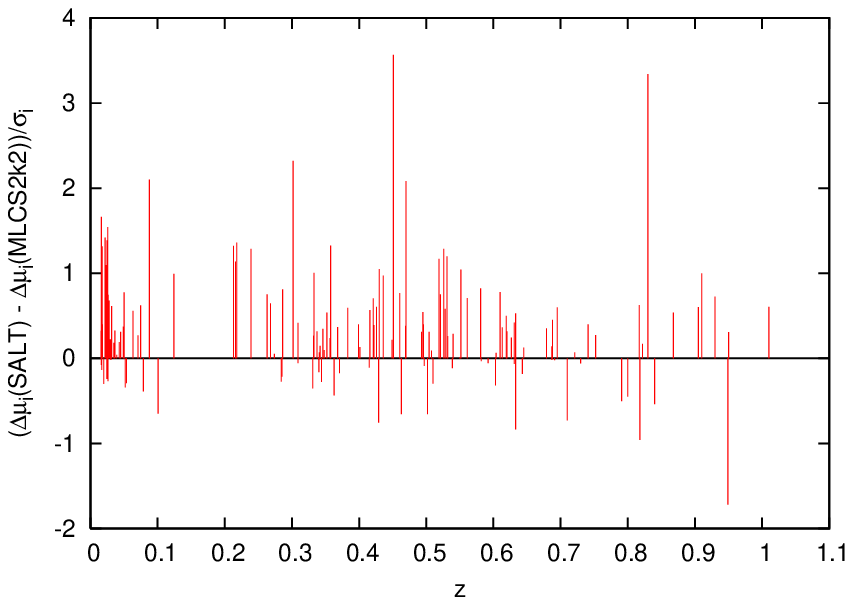}}
\caption{Differences of apparent magnitudes obtained by the SALT and the
  MLCS2k2 fitting method \subref{mudiff} and these differences devided
  by $\sigma_i$ \subref{weightmudiff}.}
\end{figure}

It is noticeable that the two light-curve fitting methods for the
ESSENCE set yield very different
results. (For a discussion on these differences at low redshifts, see
\cite{conley}.) 
This could be an indication that at least one of the two
fitting methods does not give the correct result. In order to consider
this possibility we take a look at the differences between
$\Delta\mu_i$ obtained by SALT and $\Delta\mu_i$ obtained by
MLCS2k2. They are shown in figure \ref{mudiff} for the 153 SNe
contained in both ESSENCE sets. For some SNe the difference is very
large (namely up to one magnitude). In order to see if the two sets
are consistent, we need to know how large the difference is in terms
of the error $\sigma_i=\left[\sigma^2_i\mbox{(SALT)} +
  \sigma^2_i\mbox{(MLCS2k2)}\right]^{1/2}$. The result is shown in
figure \ref{weightmudiff}.  We want the systematical error due to the
different fitting methods to be smaller than the statistical error of the
observational data.  Thus, we discard all SNe with a
difference in $\Delta\mu_i$ larger than $1\sigma_i$ and those that
are only contained in one of the two ESSENCE sets, which leaves 129 SNe
in the sets. Again we counted the SNe indicating acceleration or
deceleration for each set and each calibration method. The result is
shown in table \ref{acctab2}. The outcome of this test did not change
qualitatively by discarding suspicious SNe from the sets.
Thus, we will use all the SNe of the ESSENCE sets in the following.

\begin{table}
\begin{indented}\lineup
\caption{\label{acctab2} Number of SNe indicating acceleration or
  deceleration for SNe of the ESSENCE sets with
  $|\Delta\mu_i\mbox{(SALT)}-\Delta\mu_i\mbox{(MLCS2k2)}|/\sigma_i \le
  1$. The total number of SNe in each set is 129. Again, the most
  discrepant results are highlighted.}
\item[]\begin{tabular}{@{}lllll} \br 
&\centre{2}{ESSENCE (MLCS2k2)} &\centre{2}{ESSENCE (SALT)} \\ 
& Riess & Sandage & Riess & Sandage \\\mr 
acceleration (95\% CL) & 33 & 11 & 69 & 30 \\ 
acceleration (99\% CL) &\08 &\0{\bf 2} &{\bf 40} & 14 \\\mr 
deceleration (95\% CL) &\01 &\02 &\00 &\06 \\ 
deceleration (99\% CL) &\00 &\0{\bf 1} &\0{\bf 0} &\01 \\ \br
\end{tabular}
\end{indented}
\end{table}

The error of the distance modulus contains the peculiar velocities
$v_{\mbox{\scriptsize pec}}$ of the SNe. In the ESSENCE sets
$\sigma_v$ is assumed to be 400km/s for all SNe. We
wanted to know how sensitive our test is with respect to changes of
the peculiar velocity. Table \ref{acctabvel} shows that varying
$\sigma_v$ almost does not change the result with
the exception of ESSENCE (MLCS2k2) in the 
Sandage calibration, where the number of SNe indicating deceleration
at a 95\% CL is 7 for $\sigma_v=400$km/s and
500km/s, but 11 for $\sigma_v=300$km/s.

\begin{table}
\begin{indented}\lineup
\caption{\label{acctabvel} Number of SNe indicating acceleration or
  deceleration for SNe of the ESSENCE sets with different peculiar
  velocity dispersions $\sigma_v$.}
\item[]\begin{tabular}{@{}llllll} \br &&\centre{2}{ESSENCE (MLCS2k2)}
  &\centre{2}{ESSENCE (SALT)} \\ $\sigma_v$[km/s] &&
  Riess & Sandage & Riess & Sandage \\\mr 
  300 
  &acceleration (95\% CL) & \037 & \014 & \099 & \053 \\ 
  &acceleration (99\% CL) & \011 &\0\03 & \064 & \030 \\ 
  &deceleration (95\% CL) &\0\03 & \011 &\0\01 &\0\07 \\ 
  &deceleration (99\% CL) &\0\02 &\0\03 &\0\00 &\0\02 \\ \mr 
  400 
  &acceleration (95\% CL) & \037 & \014 & \096 & \053 \\ 
  &acceleration (99\% CL) & \011 &\0\03 & \064 & \030 \\ 
  &deceleration (95\% CL) &\0\03 &\0\07 &\0\01 &\0\07 \\ 
  &deceleration (99\% CL) &\0\02 &\0\03 &\0\00 &\0\02 \\ \mr 
  500 
  &acceleration (95\% CL) & \036 & \013 & \095 & \053 \\ 
  &acceleration (99\% CL) & \010 &\0\03 & \063 & \030 \\ 
  &deceleration (95\% CL) &\0\03 &\0\07 &\0\01 &\0\05 \\ 
  &deceleration (99\% CL) &\0\02 &\0\03 &\0\00 &\0\02 \\ \mr 
  \multicolumn{2}{@{}l}{number of SNe} & 162 & 162 & 178 & 178 \\
\br
\end{tabular}
\end{indented}
\end{table}

\subsection{Averaging over SN data}

Until now, we have only made tests for single SNe. The
problem in combining these data is that the quantity $\Delta\mu$
depends on the redshift $z$. Thus the data from different redshifts do
not have the same mean value. Nevertheless it is possible to average
over $\Delta\mu_i$. Then the result of course depends on how many SNe
from a certain redshift are used to calculate this mean
value. Therefore, the average over all SNe of a set does not
characterize the function 
$\Delta\mu(z)$. But still, when combined with its standard deviation,
the mean value can be an evidence for acceleration or
deceleration. Thus, we define
\begin{equation}
\overline{\Delta\mu} =
\frac{\sum_{i=1}^Ng_i\Delta\mu_i}{\sum_{i=1}^Ng_i} \;,
\end{equation}
where $g_i=1/\sigma_i^2$. The mean value is calculated in such a way
that data points with a small error are weighted more than those with
large errors. The standard deviation of the mean value is calculated
by
\begin{equation}
\sigma_{\overline{\Delta\mu}} =
\left[\frac{\sum_{i=1}^Ng_i\left(\Delta\mu_i -
    \overline{\Delta\mu}\right)^2}{(N-1)\sum_{i=1}^Ng_i}\right]^{\frac{1}{2}}
\;.
\end{equation}

\begin{figure}
\centering \subfloat[Riess calibration]{
  \includegraphics*[width=7.5cm]{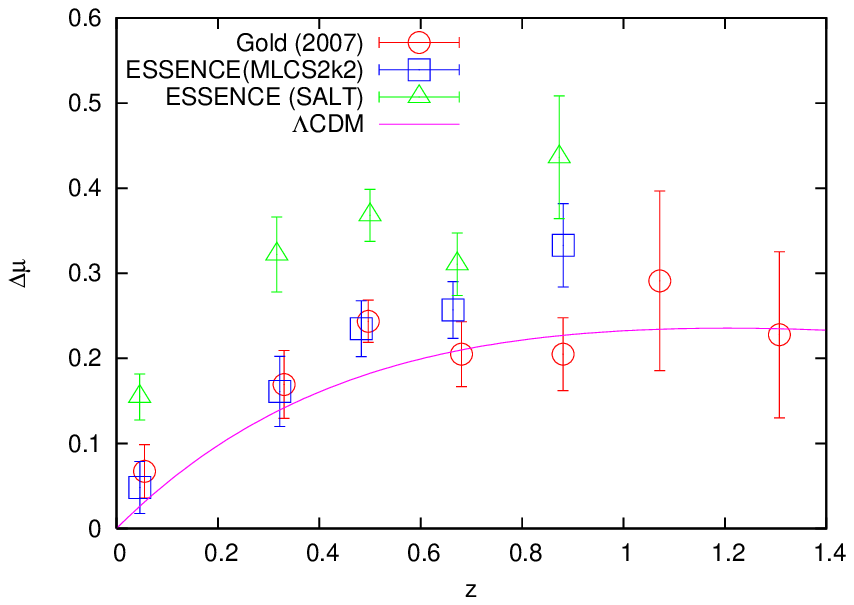}} \hfill
\subfloat[Sandage calibration]{
  \includegraphics*[width=7.5cm]{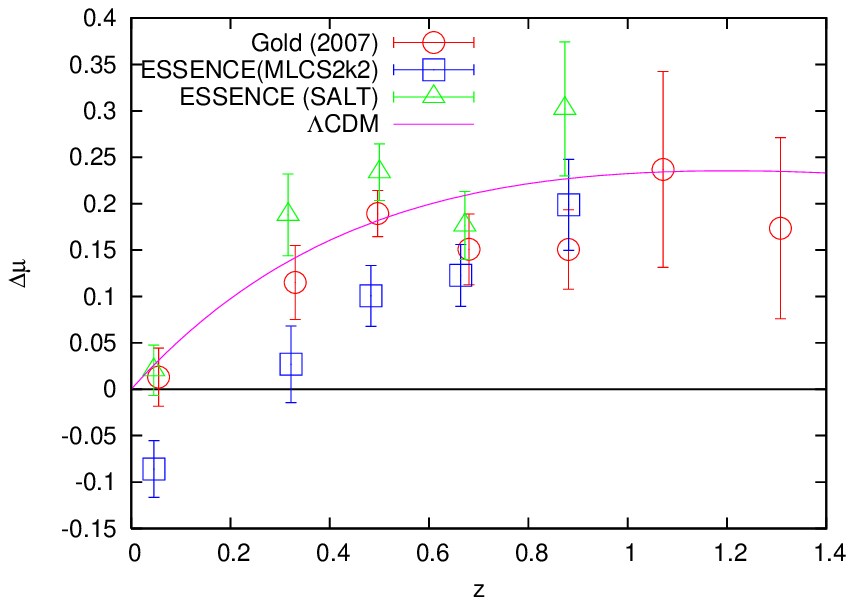}}
\caption{Magnitude $\Delta\mu$ averaged over redshift bins of width 0.2 for
  different data sets.}
\label{zbinfig}
\end{figure}

We start with averaging $\Delta\mu_i$ over redshift bins of width
0.2. For all data sets the averaged value of $\Delta\mu$ increases
with redshift (see figure \ref{zbinfig}) which could be expected for
an accelerated expansion. The curve for a flat $\Lambda$CDM with
$\Omega_m=0.3$ has its maximum
at $z=1.2$ and then decreases again. Unfortunately, there are not
enough SNe at high redshifts to state a possible decrease of
$\Delta\mu$ at some point. The differences in the data points of the
two ESSENCE sets again seem to be too large. MLCS2k2 gives smaller
values than the SALT fitter for all redshift bins. As $\Delta\mu(z)$
should be close to zero at low redshifts in a
homogeneous and isotropic universe, the MLCS2k2 data look more
consistent with our assumptions in the Riess calibration, whereas SALT
gives the better results 
in the Sandage calibration. The Gold sample has sensible values in
both calibrations.
$\overline{\Delta\mu}$ devided by the error
$\sigma_{\overline{\Delta\mu}}$ gives the evidence for acceleration in
each redshift bin. As can be seen in table \ref{zbin}, the strongest
evidence is given in redshifts between 0.4 and 0.8.

\begin{table}
\begin{indented}\lineup
\caption{\label{zbin} Statistical evidence
  $\overline{\Delta\mu}/\sigma_{\overline{\Delta\mu}}$ 
  within the given redshift range for
  a flat and an open universe.}
\item[]\begin{tabular}{@{}lllllll} \br &\centre{2}{Gold 2007 (MLCS2k2)}
  &\centre{2}{ESSENCE (MLCS2k2)} &\centre{2}{ESSENCE (SALT)} \\ 
z & Riess & Sandage & Riess & Sandage & Riess & Sandage \\ \mr 
flat universe &&&&&&\\
0.0 -- 0.2 & 2.1 & 0.4 & 1.6 &\-2.8 & \05.7 & 0.8 \\ 
0.2 -- 0.4 & 4.2 & 2.9 & 3.9 &  0.7 & \07.3 & 4.3 \\ 
0.4 -- 0.6 & 9.8 & 7.6 & 7.1 &  3.1 &  12.1 & 7.7 \\ 
0.6 -- 0.8 & 5.4 & 4.0 & 7.7 &  3.7 & \08.5 & 4.8  \\ 
0.8 -- 1.0 & 4.8 & 3.5 & 6.8 &  4.1 & \06.0 & 4.2 \\ 
1.0 -- 1.2 & 2.8 & 2.2 &&&& \\ 
1.2 -- 1.4 & 2.3 & 1.8 &&&& \\ \mr
open universe &&&&&&\\
0.0 -- 0.2 &  2.1 &  0.4 & 1.5 &\-2.8 &\05.7 & 0.7 \\
0.2 -- 0.4 &  3.5 &  2.2 & 3.2 &\-0.0 &\06.6 & 3.6 \\
0.4 -- 0.6 &  7.4 &  5.2 & 5.5 &  1.4 & 10.2 & 5.8 \\
0.6 -- 0.8 &  2.8 &  1.4 & 4.9 &  0.9 &\05.9 & 2.2 \\
0.8 -- 1.0 &  1.4 &  0.2 & 3.9 &  1.2 &\04.0 & 2.2 \\
1.0 -- 1.2 &  0.9 &  0.4 &&&&\\
1.2 -- 1.4 &\-0.2 &\-0.7\\
\br
\end{tabular}
\end{indented}
\end{table}

Next we average over all SNe of each data set with a redshift $z\ge0.2$.
We discard SNe with a smaller redshift for the following reasons: (a)
$\Delta\mu(z)$ is expected to be relatively close to zero for small
redshifts. Therefore, nearby SNe do not contribute to the evidence for
acceleration. (b) Considering only the nearby universe, local effects can
modify the results for the distance modulus as the cosmological
principal is not valid on small scales. (c)
Another disadvantage of nearby SNe is that they were
observed with many different telescopes and thus the obtained
systematic error is potentially higher.
Table \ref{meantab2} shows the mean values
$\overline{\Delta\mu}$ and their standard deviations
$\sigma_{\overline{\Delta\mu}}$
evaluated by using only SN data with $z\ge0.2$. 
$\overline{\Delta\mu}$ is positive for all data sets.
$\overline{\Delta\mu}$ divided by $\sigma_{\overline{\Delta\mu}}$
indicates the confidence level at which an accelerated expansion can
be stated. The weakest evidence for acceleration is given for ESSENCE
(MLCS2k2) in the Sandage calibration. Here the mean value lies $5.2
\sigma$ above 0, i.e. above the value for a universe that neither
accelerates nor decelerates. In the other cases the confidence level is even
larger, up to $17.0\sigma$ for ESSENCE (SALT) in the Riess
calibration.

\begin{table}
\begin{indented}\lineup
\caption{\label{meantab2} Mean values and standard deviations of
  $\Delta\mu$ obtained by using only SNe with $z\ge 0.2$ for a flat
  universe.} 
\item[]\begin{tabular}{@{}lllllll} \br &\centre{2}{Gold 2007 (MLCS2k2)}
  &\centre{2}{ESSENCE (MLCS2k2)} &\centre{2}{ESSENCE (SALT)} \\ 
& Riess & Sandage & Riess & Sandage & Riess & Sandage \\ \mr
  $\overline{z}$ & 
\00.63 & 0.63 & \00.54 & 0.54 & \00.51 & \00.51 \\ 
$\overline{\Delta\mu}$ & 
\00.2196 & 0.1655 & \00.2398 & 0.1056 & \00.3457 & \00.2115 \\ 
$\sigma_{\overline{\Delta\mu}}$ & 
\00.0167 & 0.0167 & \00.0201 & 0.0201 & \00.0203 & \00.0203  \\ 
$\overline{\Delta\mu}/\sigma_{\overline{\Delta\mu}}$ & 
13.1 & 9.9 & 11.9 & 5.2 & 17.0 & 10.4 \\ \br
\end{tabular}
\end{indented}
\end{table}

\section{Open and closed universe}\label{openclosed}
Although there are good reasons to believe that the universe is flat
we give up this assumption in the following section. In an open universe
the luminosity distance of a universe that neither accelerates nor
decelerates is given by
\begin{equation}
d_{\mbox{\scriptsize L}}(z) = \frac{1+z}{H_0\sqrt{\Omega_k}}
\sinh\left( \sqrt{\Omega_k}\ln(1+z)\right)
\end{equation}
and thus depends on the density parameter of the scalar curvature
$\Omega_k$. $d_{\mbox{\scriptsize L}}$ increases with increasing
$\Omega_k$ and thus the evidence for acceleration becomes weaker. As
we are interested in the lower limit of this evidence we have to take
the highest possible value for the scalar curvature, i.e. $\Omega_k=1$,
which corresponds to an empty universe. (Here we allow ourself to make
use of the Einstein equation.)
Then equation \eref{deltamui} for $\Delta\mu_i$ changes to
\begin{equation}
\Delta\mu_i = \mu_i - \mu(q=0) = \mu_i - 5\log\left[
  \frac{1}{H_0}(1+z_i)\sinh[\ln(1+z_i)] \right] -25 \,.
\end{equation}

\begin{table}
\begin{indented}\lineup
\caption{\label{opentab} Statistical evidence
  $\overline{\Delta\mu}/\sigma_{\overline{\Delta\mu}}$  
  for an open universe (obtained by using SNe within the redshift
  range $0.2\le z <1.2$), a flat and a closed universe ($0.2\le z$).}
\item[]\begin{tabular}{@{}lllllll} \br &\centre{2}{Gold 2007 (MLCS2k2)}
  &\centre{2}{ESSENCE (MLCS2k2)} &\centre{2}{ESSENCE (SALT)} \\ 
& Riess & Sandage & Riess & Sandage & Riess & Sandage \\ \mr
open universe & 
\08.0 & \04.9 & \08.8 & 1.8 & 13.8 & \07.2 \\
flat universe &
 13.1 & \09.9 & 11.9 & 5.2 & 17.0 & 10.4 \\
closed universe &
 13.1 & \09.9 & 11.9 & 5.2 & 17.0 & 10.4 \\\br
\end{tabular}
\end{indented}
\end{table}

The evidence for accelerated expansion is then calculated in the same
way as for a flat universe. The result obtained by averaging over
redshift bins is shown in table \ref{zbin}.
For the overall average we only used SNe between
redshift 0.2 and 1.2 because including higher redshifts would weaken
the evidence. This is due to the fact that the values
of $\Delta\mu_i$ become negative when the phase of deceleration within
the redshift range over which is integrated is large enough. The
result is shown in table \ref{opentab}. As could be expected, the
evidence is now much weaker than for a flat universe. But we still
find a hint of acceleration at 1.8$\sigma$.

For a closed universe we have a different situation: Here
$d_{\mbox{\scriptsize L}}$ decreases with increasing spatial
curvature. Thus the lower limit of the evidence for acceleration is
given in the case of the lowest possible curvature which corresponds
to a flat universe.

\section{Conclusion}\label{concl}
We tested the cosmic expansion without specifying any density
parameters or parameterizing kinematical quantities. This was not
possible without assuming a certain calibration of $M$ and
$H_0$. Therefore, we considered two very different calibrations. We
also considered two different data sets and two different light curve
fitters and varied the peculiar velocity dispersion. Using single SNe
for the test has already given a 
clear indication of acceleration in a flat universe although a few SNe
strongly favour deceleration. We find large systematic effects already
at the level of individual SNe.

Similar analyses have already been done by other groups \cite{santos,
  gong}. But note that the work of Santos et al.~\cite{santos}
exhibits two major shortcomings. First, they did not calibrate the SN
data consistently. They took a certain value for $H_0$ but kept the
arbitrary value of the absolute magnitude $M$ given in the set which
led to distance moduli that are too high. Thus, they concluded
erroneously that there was a recent phase of super-acceleration. The
second problem is that they did not realize that due to the integration
over redshift this method is not suitable to determine the transition
redshift $z_{\mbox{\scriptsize t}}$ between deceleration and
acceleration. Thus, their value of $z_{\mbox{\scriptsize 
t}}$ does not hold. The shortcoming of the work of Gong et
al.~\cite{gong} is that they have not applied any statistics at all.

Instead of only considering single SNe, we obtained a more significant
result by averaging over the SN
data of each set. Here it is justified to discard nearby SNe in order
to decrease systematics from using different telescopes or from local
effects. We argue that we reject the null hypothesis of no
acceleration for a spatially flat, homogeneous and isotropic universe
at high confidence ($>5\sigma$).
However, the results show large differences for each data set, calibration
and light-curve fitter. E.g. changing the calibration from Sandage to
Riess calibration in a flat universe using ESSENCE(MLCS2k2) increases
the evidence for 
acceleration from 5.2$\sigma$ to 11.9$\sigma$. For the two different
light-curve fitters applied to the ESSENCE SNe using the Riess
calibration, we get values of 11.9$\sigma$ and 17.0$\sigma$.  

Wood-Vasey et al.~argue that the differences
due to different light curve fitters are not important as they
disappear when marginalization is applied \cite{essence}. This is 
true if the data are only used to determine the parameter values of a
certain cosmological model as in this case no calibration of $M$ and
$H_0$ is needed. But nevertheless, the fitters should give the same
result for a given absolute magnitude $M$. If this is not the case,
a systematic error in the determination of the apparent magnitude
$m$ is introduced by at least one of the fitters. Systematics due to
different fitters have also been described in \cite{conley} and those
within the Gold sample in \cite{jassal1,jassal2,nesseris}.

The different results obtained by the two calibrations are very
surprising. Our test only depends on the reduced distance modulus
$\mathcal{M}=M-5\log H_0+25$ and not on the absolute magnitude or the
Hubble constant individually. (Note that this is the quantity which is
marginalized when fitting cosmological parameters.) 
Starting from different calibrations $M_1$ and $M_2$ leads to
different values of $H_0$, 
determined by observation of SNe, in such a way that $\mathcal{M}$
should in principle be the same for both values of $M$.
Thus our test should not depend on the calibration. However, the fact
that the calibration changes our result significantly can be explained
as follows: $H_0$ is determined by observing SNe. Riess et al.~and
Sandage et al.~use different sets of SNe and different fitting methods
for this determination. The systematic errors due to different sets and
fitters then of course 
also influence the result for the Hubble constant leading to different
values $\mathcal{M}_1$ and $\mathcal{M}_2$.

For a conservative conclusion, we need to take the set that gives the
weakest evidence for acceleration, namely ESSENCE (MLCS2k2) in the
Sandage calibration. Using this set, accelerated expansion can be stated
at 5$\sigma$ if we assume a spatially flat or closed universe. In an
open universe the evidence is much weaker, namely 1.8$\sigma$. These
results only hold if the correct analysis of SNe is somewhere in the
range of the cases we considered here. But as we observe enormous
systematic effects in our test, we cannot be sure of this assumption.
Thus, it is a major issue to better
understand how SNe have to be observed and analyzed. 

Remember that results of this work are only valid if the universe is
homogeneous and isotropic. The situation changes dramatically if we
give up those assumptions. There exist enormously large structures in
the universe, an example is the Sloan Great Wall with an extension of
$\sim$400 Mpc \cite{gott}. Besides, there are also big voids and
superclusters on a 100 Mpc scale. This roughness of the universe could
spoil the validity of equation (1). 
Indeed it is possible to construct
inhomogeneous models that can describe the observational data without
the need of an average acceleration \cite{celerier,havard,enqvist,ishak}.

Some observational evidence for inhomogeneity and anisotropy in SN
Hubble diagrams has recently been presented by \cite{dyer,weinhorst},
probably due to large scale bulk motion and perhaps systematic
effects. A non-trivial scale dependence of the Hubble rate due to the
so-called cosmological backreaction
\cite{buchert,schwarz,rasanen,kolb,vanderveld,wiltshire}  
has been shown in \cite{li}. This effect of averaging can also mimic
curvature effects 
\cite{li1} and thus strongly influence the reconstruction of the dark
energy equation of state, as has been shown in \cite{bassett}.

Thus we think a major task must be to establish the acceleration of
the universe independently of the assumption of strict homogeneity and
isotropy. 

\ack
We thank Dragan Huterer, Thanu Padmanabhan, Aleksandar Raki\'{c}, Adam
Riess and Bastian Weinhorst for discussions, comments and references
to the literature. This work is supported by the DFG under grant GRK 881.

\end{document}